\documentclass[12pt,preprint,flushrt]{aastex}

\newcommand{\e}            {\mbox{$^{-1}$}}
\newcommand{\ee}           {\mbox{$^{-2}$}}

\newcommand{\simgt}        {\gtrsim}

\newcommand{\pp}           {\noindent\hangindent 20pt\hangafter=1}

\def\startfigcap{\vspace*{2.0\baselineskip}\bgroup\leftskip 0.45in\rightskip 0.45in}
\def\endfigcap{\par\egroup\vspace*{2.0\baselineskip}}
\def\plotfiddle#1#2#3#4#5#6#7{\centering \leavevmode
\vbox to#2{\rule{0pt}{#2}}
\includegraphics{#1}}

\begin{document}

\title{Detection of cool dust around the G2V star HD~107146}
\author{Jonathan P. Williams\altaffilmark{1},
Joan Najita\altaffilmark{2},
Michael C. Liu\altaffilmark{1,6},
Sandrine Bottinelli\altaffilmark{1},
John M. Carpenter\altaffilmark{3},
Lynne A. Hillenbrand\altaffilmark{3},
Michael R. Meyer\altaffilmark{4}
and David R. Soderblom\altaffilmark{5}}

\shorttitle{Cool dust around a G2 star}
\shortauthors{Williams et al.}

\begin{abstract}
We report the detection of dust emission at sub-millimeter wavelengths from
HD~107146, a G2V star with an age estimated to lie between 80 and 200~Myr.
The emission is resolved
at 450~\micron\ with a size $300~{\rm AU}\times 210~{\rm AU}$.
A fit to the spectral energy distribution gives a dust temperature of
51~K and dust mass of $0.10~M_\oplus$. No excess emission above the
photosphere was detected at 18~\micron\ showing that there is very
little warm dust and implying the presence of a large inner hole,
at least 31~AU ($\sim 1''$) in radius, around the star.
The properties of this star-disk system are compared with similar
observations of other systems.  We also discuss prospects for future 
observations that may be able to determine whether the inner
hole is maintained by the dynamical effect of an unseen orbiting
companion.
\end{abstract}
\keywords{circumstellar matter --- planetary systems --- stars:individual(HD 107146)}

\altaffiltext{1}{Institute for Astronomy, 2680 Woodlawn Drive,
Honolulu, HI 96822; jpw, mliu, sandrine@ifa.hawaii.edu}
\altaffiltext{2}{National Optical Astronomy Observatory,
950 North Cherry Avenue, Tucson, AZ 85719; najita@noao.edu}
\altaffiltext{3}{Department of Astronomy/Astrophysics, MS 105-24, 
California Institute
of Technology, 1201 East California Boulevard, Pasadena, CA 91125;
jmc, lah@astro.caltech.edu}
\altaffiltext{4}{Steward Observatory, University of Arizona,
933 North Cherry Avenue, Tucson, AZ 85721; mmeyer@as.arizona.edu}
\altaffiltext{5}{Space Telescope Science Institute, 3700 San Martin Drive,
Baltimore, MD 21218; drs@stsci.edu}
\altaffiltext{6}{Hubble Fellow}

\section{Introduction}
Protostellar disks dissipate over a period of several Myr
as their constituent dust and gas either accretes onto the star,
is dispersed by processes such as stellar winds and photoevaporation,
or aggregates into planetesimals (Hollenbach, Yorke, \& Johnstone 2000).
Possible observational examples of disk systems in the process of 
dissipating are the ``transitional disks'', a thus far small class of 
objects which includes sources such as V819 Tau (Skrutskie et al.\ 1990), 
HR4796A (Jura et al.\ 1993; Jayawardhana et al.\ 1998; 
Koerner et al.\ 1998; Schneider et al.\ 1999; Telesco et al.\ 2000) 
and HD141569A (Weinberger et al.\ 1999; Augereau et al.\ 1999; 
Fisher et al.\ 2000).  
 
Dramatic evidence for disk dispersal through planetesimal formation
is the detection of numerous planets, and some planetary systems,
around nearby stars (Marcy, Cochran, \& Mayor 2000).
Planetesimals are believed to grow into planets through collisional 
agglomeration. But collisions between planetesimals are also 
expected to create a cascade of smaller particles, and in this way 
a relatively old circumstellar disk may regenerate its own dust.

Such second generation dust was first detected around main-sequence
stars by IRAS (Aumann et al. 1984).  The IRAS results indicated 
that $\sim 15$\% of main-sequence stars possess dusty disks 
(Lagrange, Backman, \& Artymowicz 2000), the brightest of 
which have been detected at sub-millimeter wavelengths 
(Zuckerman \& Becklin 1993; Greaves et al.\ 1998; 
Sylvester, Dunkin, \& Barlow 2001; 
see also Wyatt, Dent, \& Greaves 2003). 
Sub-millimeter observations are useful because 
they place strong constraints on the dust mass of the disk, 
due to the low optical depth of dust grains in the sub-millimeter  
compared to their opacities at shorter wavelengths. 
Because of their small angular size and weak emission,
only four debris disk systems have been spatially resolved at
these wavelengths to date (Holland et al. 1998; Greaves et al. 1998).
The maps of these systems show interesting asymmetries that have been 
interpreted as the dynamical signature of a planetary companion 
(Holland et al. 2003; Wilner et al. 2002; Greaves et al. 1998).

Although the majority of the well-studied debris 
(and transitional) disk systems surround early-type stars 
(Vega, Fomalhaut, $\beta$ Pic, HR4796A, and HD141569A are all 
A stars), the expectation is that the process of disk dissipation 
and the regeneration of dust through planetesimal collisions is 
also a part of the evolutionary history of lower mass stars 
and, indeed, our own solar system.  Thus, it is of considerable 
interest to identify debris disks associated with young solar-mass 
stars, since the detailed study of these objects can provide 
direct insight into the evolutionary history of our solar 
system.  The $\epsilon$ Eri system ($d=3.2$ pc) is one exciting example 
(Greaves et al.\ 1998; Hatzes et al.\ 2000). 
In this {\it Letter}, we present sub-millimeter observations of a 
dusty disk around a close solar analogue, 
the nearby young G2V star HD~107146.

HD~107146 was first identified as an ``excess dwarf''
on the basis of IRAS colors by Silverstone (2000).
We observed it as part of a program to provide ground-based
support for the ``Formation and Evolution of Planetary Systems''
SIRTF Legacy Project\footnotemark[7]\footnotetext[7]{Information
concerning this program can be found at http://feps.as.arizona.edu}.
HD~107146 was selected for inclusion in the Legacy project based on its
distance (28.5~pc; Perryman et al. 1997) and 
its high Ca II H and K index $\log R^\prime_{\rm HK}=-4.28$,
which is indicative of youth.
An age range can be estimated from its lithium equivalent width,
125~m\AA\ (Wichmann, Schmitt, \& Hubrig 2003),
which places it at the lower envelope of the 125~Myr Pleiades distribution
(Soderblom et al. 1993) but above the distribution for the 250~Myr old M34
(Jones et al. 1997) at the effective temperature of a G2V star.
Wichmann et al. also show that its space motions are similar to the
Pleiades moving group.
Its x-ray luminosity, $L_X=2\times 10^{29}$~erg~s\e\ (Voges et al. 1999),
is again similar to solar-type Pleiads (Micela et al. 1999)
but weaker than that of the average solar-type stars in the
80~Myr old Alpha Per cluster (Prosser et al. 1996).
Together, these indicators suggest an age for HD~107146 between 80 and 200~Myr.
Its location in the HR diagram is also consistent with this age range
but allows the possibility that it may be as young as 30~Myr if the star 
has just reached the main sequence.

Our observations, made at sub-millimeter and mid-infrared wavelengths, 
are detailed in \S 2. The results from the imaging and fits to the
spectral energy distribution (SED) are presented in \S 3.
The SED shows clear evidence for a large inner hole, although 
the hole is too small to be resolved in the sub-millimeter maps 
presented here.
In \S4, we compare the properties
of the HD107146 star-disk system to other debris and transitional 
disk systems. We also discuss prospects for future 
observations that may be able to determine whether the inner
hole is maintained by the dynamical effect of an unseen orbiting
companion.

\section{Observations}
HD~107146 was mapped using the SCUBA bolometer array
at the James Clerk Maxwell Telescope (JCMT) on Mauna Kea, Hawaii in dry,
stable conditions during February 17-19, 2003.
Precipitable water vapor levels were less than 0.5 to 1~mm during
the observations and zenith optical depths ranged from 0.1 to 0.18 at
850~\micron\ and 0.4 to 0.85 at 450~\micron.
Observations were made simultaneously at 450 and 850~\micron\ using a 64 point
``jiggle'' pattern to produce Nyquist sampled images at each wavelength.
Two maps were rejected in the reduction process due to an error in the
telescope tracking software at high elevation (Coulson, personal communication).
The final maps presented here are the median of 23 maps made at a range
of image rotation angles on the array rebinned to a $1''$ rectangular grid
in equatorial coordinates. Pointing was checked between
individual maps using Virgo A and 3C273 and was accurate to an rms
error of $2\farcs 5$. Calibration was carried out by observing the CRL 618
planetary nebula and bootstrapped to observations of Mars later in the
night. Based on the gain variations from night to night, the calibration
accuracy is estimated to be 20\% at 850~\micron\ and 30\% at 450~\micron.
The total on source integration time was 4.0 hours and the noise in
the final maps was 4 and 13~mJy~beam\e\ at 850 and 450~\micron\
respectively. The resulting peak signal-to-noise ratio is $\sim 7$
in both maps but the calibration uncertainty dominates the error in
the absolute flux measurements.

We obtained mid-IR photometry of HD~107146 from the Keck~II Telescope on
Mauna Kea, Hawaii using the facility instrument LWS.
We observed the star on February 19 and 20, 2003~UT using filters centered
at 11.7~\micron\ (10.5--12.9~\micron) and 17.8~\micron\
(17.3--18.2~\micron).  Contemporaneously and at similar airmasses, we
observed the bright standard stars $\alpha$~CrB and $\mu$~UMa from
Tokunaga (1988) for photometric calibration.  Both nights were
characterized by dry conditions with an estimated 0.5--1.5~mm of
precipitable water, as determined from sky dip measurements from the
JCMT.  Seeing conditions were poor and variable,
leading to mid-IR images with 0.4\arcsec\ to 0.6\arcsec\ FWHM.
Observations were conducted in the standard ``chop/nod'' mode,
which involves switching between three sky positions using fast chopping
of the secondary mirror and slower nodding of the telescope itself. This
allows for effective subtraction of the very bright and variable thermal
emission from the sky and telescope in the reduction process.

HD~107146 was well-detected in both filters, with a formal signal-to-noise
ratio $>$15 for the 17.6~\micron\ data and $>$50 for the 11.7~\micron\ data.
However, the seeing conditions limit the photometric precision, due to varying
image quality from the science target to the standard stars.  These
conditions also prevented any useful constraints on any extended mid-IR
emission from HD~107146.  Photometric errors were first determined from
the scatter in measurements done with apertures scaled by the FWHM of
the images.  We then added in quadrature an error term to account for
seeing mismatch (10~mJy at 11.7~\micron, 8~mJy at 17.6~\micron),
estimated from analyzing curve of growth photometry for standard stars
observed during the course of the entire night. The absolute flux
calibration is based on the flux of Vega ($\alpha$~Lyr) compiled by
Tokunaga (1988). The LWS and SCUBA fluxes are listed in Table~1.

\section{Results}
The resulting maps of the emission at 450 and 850~\micron\ 
are shown in Figure~1.
There is a $2\farcs 9$ offset between the two maps, as measured by gaussian
fits to the data clipped at the FWHM level.
The centroid of the emission is consistent with the stellar position 
to within $4\farcs 4$ (450~\micron) and $1\farcs 6$ (850~\micron). 
It is not clear how significant are such small offsets given the 
relatively low signal-to-noise in the data and the possibility of 
systematic pointing errors as large as $2\farcs 5$.
No other source within this angular range of the star was apparent
in the Digital Sky Survey, 2MASS images,
the Keck imaging at 11.7 and 17.8~\micron, and the FIRST 20~cm survey.
In addition, Blain et al. (1999) predict $100-300$ sources deg\ee\
with fluxes greater than 20~mJy at 850~\micron, implying a probability
less than $1.4\times 10^{-3}$ that an unrelated background object this
bright would be found within $4\farcs 4$ of the star.
Since the probability of unrelated submillimeter emission is very low,
we assume that the SCUBA source is a disk associated with the star and
discuss its properties in this context.

Allowing for pointing errors, the resolutions of the maps are
$14\farcs 5$ at 850~\micron\ and $8''$ at 450~\micron.
The gaussian fits give sizes of $15\farcs 1\times 14\farcs 9$
and $13\farcs 2\times 10\farcs 9$ (position angle $-35^\circ$) respectively.
The elongation at 450~\micron\ is apparent at the half-power level
and is therefore not due to the JCMT beam pattern which shows
significant non-circularity only at the 10\% level.
The slight extension in the 450~\micron\ map is also seen in
crosscuts through the image along the major axis. To increase
the signal-to-noise ratio, an average crosscut over a range of
position angles $-35^\circ\pm 10^\circ$ centered on the peak of emission
is shown for each wavelength in the lower panels of Figure~1.
The same crosscut averaging, shown as dashed lines, was performed for a
map of Mars, taken from the last night of observations,
February $19^{\rm th}$, when the Martian diameter was $5\farcs 7$.
We conclude that the disk around HD~107146 is marginally
resolved at the $8''$ resolution of the 450~\micron\ data.
Subtracting the beam size in quadrature from the gaussian fit
gives an angular size of $10\farcs 5\times 7\farcs 4$ for the disk
corresponding to $300~{\rm AU}\times 210~{\rm AU}$.

The fluxes measured from the LWS and SCUBA observations are listed
in Table~1. The SED of the source,
from optical to sub-millimeter wavelengths, is plotted in Figure~2.
The {\it UBVRI} photometry is from Landolt (1983), the near-infrared
fluxes in the {\it JHK} bands are from 2MASS, and the 10~\micron\ point is
from Palomar observations (Metchev, Hillenbrand, \& Meyer 2003).
IRAS fluxes were determined by color correcting the quoted values
in the Faint Source Catalog. The IRAS 12~\micron\ flux is consistent
with the Keck 11.7~\micron\ observation and is not plotted.
The stellar photosphere was fit by a Kurucz model
($T_{\rm eff}=5750~{\rm K}, \log g=4.5$, solar metallicity)
with a power law extrapolation beyond 10~\micron.
Strong excess emission is apparent beyond 25~\micron.
In order to compare the properties of the excess with that detected
from other debris disk systems, we fit the disk SED using a 
single temperature modified black body with emission efficiency,
$Q_\lambda=1-{\rm exp}[-(\lambda_0/\lambda)^\beta]$, which has
asymptotic behavior, $Q_\lambda=1$ for $\lambda\ll\lambda_0$ and
$Q_\lambda=(\lambda_0/\lambda)^\beta$ for $\lambda\gg\lambda_0.$
The critical wavelength, $\lambda_0$, was set to 100~\micron\
for consistency with the assumptions made in previous analysis of 
debris disks SEDs (Dent et al. 2000; Wyatt, Dent \& Greaves 2003).

The parameters of the modified black body fit were measured using 
a least squares fit to the data. Errors were estimated via fits
to multiple simulations of the dust excess SED. Simulated data points
were drawn from a gaussian distribution with the mean and standard
deviation as determined by the observed data. The resulting distribution
of best fit parameter values are $T=51\pm 4$~K, $\beta=0.69\pm 0.15$.
Similarly low values of $\beta$ are found for 
other disks around main-sequence stars (Dent et al. 2000).
The dust mass, based on the fitted flux at 850~\micron\ and assuming a
dust mass absorption coefficient, $\kappa_{850}=1.7$~cm$^2$~g\e, is
$M_{\rm d}=0.10\pm 0.02~M_\oplus$. This value of $\kappa_{850}$ is chosen
for consistency with Holland et al. (1998) and Greaves et al. (1998),
but is on the high end of calculated values (Pollack et al. 1994).
For the full range of $\kappa_{850}=0.4-1.7$~cm$^2$~g\e\ discussed
in Pollack et al., the corresponding mass range is
$M_{\rm d}=0.10-0.43~M_\oplus$.
As with all sub-millimeter observations, the inferred dust masses 
do not include a potentially dominant mass component that resides 
in larger bodies (grains and planetesimals) subtending a negligible 
solid angle.  The mean parameter fit is shown in Figure~2.

The single temperature fit is a simplification that is warranted
by the small number of data points.  Nevertheless, the large dip 
in the SED at 25~\micron\ imposes a strong limit on the mass of 
warmer dust that may be present in the system.
To illustrate this constraint, a $T=100$~K, $\beta=0.7$ modified 
black body component was added to the SED fit and increased 
until the IRAS 25~\micron\ upper limit was exceeded.
This maximum allowable contribution, which corresponds to a mass limit
of $M_{\rm d}(T=100~{\rm K})=7\times 10^{-4}~M_\oplus,$ is shown in
Figure~2 but was not included in the overall fit that is shown.
The Keck 18\micron\ measurement tightly constrains the presence
of still warmer dust to a $3\sigma$ limit of 
$M_{\rm d}(T=200~{\rm K})<3\times 10^{-5}~M_\oplus$.

The limits on dust cooler than 50~K are less stringent since such 
dust emits less per unit mass and does so at longer wavelengths.
The minimum grain temperature is 23~K for blackbody particles
at the measured outer radius of the disk, 150~AU from the star.
In practice, the effect of a range of cool dust temperatures is
indistinguishable from changes in the wavelength dependence of 
the grain emissivity, parameterized by $\beta$.
However, without changing the parameters of the fit in Figure~2,
$0.13~M_\oplus$ of 23~K dust can be added before the 850~\micron\
$3\sigma$ upper limit is exceeded.
Thus, a significant fraction of the total dust mass may reside in 
cooler dust.

\section{Discussion and Summary}
We have detected strong sub-millimeter excess emission from HD~107146,
a G2V star with an age estimated to lie in the range $80-200$~Myr.
Based on a fit to the SED of the system, the mass of the emitting 
dust is estimated to be $0.10~M_\oplus$ or larger.  
We also find that the disk is marginally resolved at 450~\micron.  
The constraints placed by these observations on the mass, temperature, 
and physical extent of the dust in the system are compared in Table 2  
with the properties of other well-studied debris disk systems. 

Four of the six stars in Table~2 are A stars, a probable bias due to
the relatively high luminosity of these systems at far-infrared and
sub-millimeter wavelengths. Nevertheless, despite the range in
central star masses and luminosities, the disks have similar
properties and follow several trends.
The HD~107146 disk is quite massive, comparable in mass to the 
$\beta$~Pic disk.  Although it lies noticeably above the trend of 
decreasing mass with age that is defined by HR4796A, $\beta$ Pic, 
Fomalhaut, and Vega (Holland et al.\ 1998), it is within the scatter 
in the mass-age relation found for larger samples of dust disks  
(Wyatt et al.\ 2003).
The fractional dust luminosity ($L_d/L_*$) of HD107146 is also large; 
it is half that of $\beta$ Pic, $>10$ times that of Fomalhaut,
and is consistent with the trend of
decreasing $L_d/L_*$ with age (Spangler et al.\ 2001).
Table 2 also indicates an apparent trend of decreasing 
outer disk radius with age.  Perhaps this trend, admittedly of 
low statistical significance at present, will be verified 
when a larger number of debris disk systems have been spatially 
resolved. 

We also find that the local minimum in the SED at 25~\micron\ 
places a strong limit on the amount of warm dust in the HD~107146 
disk, $M_{\rm d}({\rm T=51 K})/M_{\rm d}({\rm T=100 K})\simgt 140$.
This lack of warm dust implies that the disk does not extend
all the way to the star.
The dust temperature, and therefore the size of this inner hole,
depends on the grain size distribution and optical properties.
For grains that emit as blackbodies at 850~\micron,
51~K dust would lie at 31~AU (Wyatt et al. 1999).
This is a lower limit to the inner radius as smaller grains could
achieve this temperature at greater distances from the star.
The other disks listed in Table~2 have similar size inner holes, 
as determined from spatially resolved images in thermal emission 
or scattered light, and are comparable in size to the Kuiper belt 
in our solar system.

Inner holes are not, by themselves, long lasting because Poynting-Roberston
drag will cause dust to spiral in from the outer disk onto the central star
on $\sim 10$~Myr timescales. Such inward migration would be evident in the
SED by the presence of warm dust (Jura et al. 1998).
Thus, the existence of inner holes has been explained as a 
consequence of either the dynamical sweeping of an orbiting 
companion (e.g., Vega; Wilner et al. 2002) or the sublimation of 
icy grains (e.g., HR~4796A; Jura et al. 1998). 
Since water ice sublimates at temperatures greater than 100~K 
(Pollack et al. 1994), this explanation can not apply to the 
HD~107146 disk where the shape of the SED places a strong limit 
on the mass of such warm dust.  The alternative explanation, 
that the inner hole is dynamically maintained by a closely 
orbiting companion, would require a fairly low-mass companion. 
The best current limit on the existence of close companions is
from Palomar adaptive optics imaging by Metchev \& Hillenbrand (2002)
who found no companions at detection limits of 11.2, 11.7, and 15.2
in absolute K band magnitude at angular separations
of $0\farcs 5, 1''$, and $2''$ respectively.
The corresponding mass limits for an age of 100 Myr are
approximately 30, 25, and $10~M_{\rm J}$ (Burrows et al. 1997).

While it has been well-recognized that the dynamical sculpting of 
disks by orbiting companions can produce inner holes and ring-like 
structures, recent studies have shown that the migration of dust in the 
presence of a residual gas disk can also induce ring-like 
structures in the dust distribution. 
For example, Takeuchi \& Artymowicz (2001)  
have shown that the dust structures seen in the HR4796A 
and HD141569A systems are qualitatively similar to those 
expected to result from the coupling between gas and dust in disks. 
However, orbiting companions may also induce significant 
departures from axisymmetry in the dust distribution 
(e.g., Liou \& Zook 1999; Ozernoy et al.\ 2000; 
Quillen \& Thorndike 2002; Moro-Martin \& Malhotra 2002), 
whereas such non-axisymmetric structures cannot be produced by 
dust migration. 
Thus, demonstrating the existence of asymmetries in the dust 
distribution, either in thermal emission or scattered light, 
as well as measuring the gas content of the HD107146 disk, is 
needed to distinguish between these two possibilities. 

At a distance of 28.5~pc, HD~107146 is relatively nearby.  
Compared to the other sources listed in Table~2, it is 
more distant than the debris disk systems that have 
been spatially resolved at submillimeter wavelengths 
($\beta$ Pic, Vega, Fomalhaut, and $\epsilon$ Eri), but it is 
at half the distance of HR4796A.  Since the HD~107146 disk is just 
resolved at the shortest operating wavelength of SCUBA, and the 
submillimeter excess is relatively bright, the Submillimeter Array 
should be able to map the morphology of the emitting dust in 
greater detail.  Phase referencing will provide a more accurate 
absolute position, and such observations could determine 
whether the dust is distributed axisymmetrically (e.g., in a ring) 
or in a more asymmetric distribution.   
For example, these observations may confirm the marginal evidence 
for an offset between the stellar position and the peak of the 
$450\micron$ emission found in the observations presented here (Figure~1). 
The offset, if real, may result from an asymmetric dust distribution, 
as has been found for several of the other debris disk systems 
in Table 2. 

The dust disk asymmetry can also be addressed by imaging the disk
in scattered light. The possibility of detecting scattered light 
from Vega-like stars is of great interest given the few such systems 
that have been detected thus far, despite extensive deep surveys 
(e.g., Kalas \& Jewitt 1996). Using the quantities listed in Table~2,
we can make a rough estimate of the relative strengths of the scattered
light from each disk. If we ignore the disk inclination and assume
a similar dust grain size distribution and albedo, the scattering
area will be proprtional to $M_d$ and the flux of scattered light
will be roughly proportional to $M_d L_*/R_{\rm out}^2d^2$, where
$d$ is the distance to the star and $R_{\rm out}$ is the physical
extent of the disk.
The factor of $R_{\rm out}^2$ arises in this expression because grains
of a given size located farther from the star intercept less starlight.
Hence the surface brightness of the disk will be proportional
to $M_d L_*/R_{\rm out}^4$.
A potentially more relevant comparison is of the contrast between
the expected surface brightness from the disk and the flux from the star,  
which is proportional to $M_d d^2/R_{\rm out}^4$ (cf. Jura et al. 1998). 
For the disk around HD107146, the expected surface brightness of the 
scattered light is one half that of the disk around $\beta$ Pic
and the expected contrast is nine times larger, suggesting that the
scattered light from HD107146 may not only be quite bright but may
also stand out against the glare of the star.
Given the low dust temperature deduced for the HD107146
disk ($<100$K), the grains are expected to be icy with a high albedo,
which favors the detection of scattered light.

Nevertheless, there are significant uncertainties associated
with this estimate. Detecting reflected light will be more challenging
if the disk is face-on rather than edge-on. In addition, the sub-millimeter
measurements from which $M_d$ is derived are primarily sensitive
to large grains ($\sim 100~\micron$), whereas the scattered light
observations will be particularly sensitive to much smaller grain
sizes. Thus, scattered light from the HD107146 disk will be weaker
if the grain size distribution is significantly skewed to large
grain sizes. Conversely, scattered light measurements can help to
constrain the grain size distribution of the disk
(e.g., Artymowicz, Burrows, \& Paresce 1989).

Future spectroscopic observations of this disk with SIRTF, particularly
in the $30-40$~\micron\ range, will place additional constraints on the
grain composition and size distribution (e.g. Wolf \& Hillenbrand 2003).
SIRTF is also expected to discover many more disks over a large range of
ages and stellar masses with consequent improvements for our understanding
of the formation and evolution of planetary systems.

\acknowledgments
We thank Anneila Sargent for counseling during an extended review process,
Herv\'e Aussel, Remo Tilanus, Iain Coulson, and Randy Campbell for
advice on instrumentation, Eric Mamajek for helpful discussions on the
age of HD~107146, and Alan Tokunaga for making the Keck time available.
We acknowledge support from NSF grant AST-0324328 (JPW),
the Beatrice Watson Parrent Fellowship at the University of Hawaii
and NASA grant HST-HF-01152.01 (MCL), and NASA contract 1224768
administered through JPL (MRM, JMC, LAH).
This research has made use of the SIMBAD database.

\section{References}
\parskip=0pt
\bigskip

\pp Artymowicz, P., Burrows, C., \& Paresce, F. 1989, ApJ, 337, 494
 
\pp Augereau, J. C., Lagrange, A. M., Moillet, D., \& Menard, F. 1999, 
	A\&A, 350, L51

\pp Aumann, H. H., Beichman, C. A., Gillett, F. C., de Jong, T.,
    Houck, J. R., Low, F. J., Neugebauer, G., Walker, R. G.,
    \& Wesselius, P. R. 1984, ApJ, 278, L23

\pp Blain, A. W., Kneib, J.-P., Ivison, R. J., \& Smail, Ian
    1999, ApJ, 512, L87

\pp Burrows, A., Marley, M., Hubbard, W. B., Lunine, J. I., Guillot, T.,
    Saumon, D., Freedman, R., Sudarsky, D., \& Sharp, C. 1997, ApJ, 491, 856

\pp Dent, W. R. F., Walker, H. J., Holland, W. S., \& Greaves, J. S.
    2000, MNRAS, 314, 702

\pp Greaves, J. S. et al. 1998, ApJ, 506, L133

\pp Greaves, J. S., Mannings, V., \& Holland, W. S. 2000, Icarus, 143, 155

\pp Fisher, R. S., Telesco, C. M., Pina, R. K., Knacke, R. F., \& 
	Wyatt, M. C. 2000, ApJ, 532, L141

\pp Hatzes, A.P., et al.\ 2000, ApJ, 544, L145

\pp Holland, W. S. et al. 1998, Nature, 392, 788

\pp Hollenbach, D. J., Yorke, H. W., \& Johnstone, D. 2000,
    in ``Protostars and Planets IV'',
    ed. V. Mannings, A.P. Boss \& S.S. Russell
    (Tucson: University of Arizona Press), 401

\pp Jayawardhana, R. Fisher, S., Hartmann, L., Telesco, C., Pina, R.,
    \& Fazio, G. 1998, ApJ, 503, L79

\pp Jones, B. F., Fischer, D., Shetrone, M., \& Soderblom, D. R.
    1997, AJ, 114, 352

\pp Jura, M., Zuckerman, B., Becklin, E. E., Smith, R. C. 1993, ApJ, 
    418, L37

\pp Jura, M., Malkan, M., White, R., Telesco, C., Pina, R., \& Fisher, R. S.
    1998, ApJ, 505, 897

\pp Kalas, P., \& Jewitt, D., 1996, AJ, 111, 1347

\pp Koerner, D. W., Ressler, M. E., Werner, M. W., \& Backman, D. E. 
    1998, ApJ, 503, L83

\pp Lagage, P. O., \& Pantin, E. 1994, Nature, 369, 628

\pp Lagrange, A.-M., Backman, D. E., \& Artymowicz, P. 2000, 
    in ``Protostars and Planets IV'',
    ed. V. Mannings, A.P. Boss \& S.S. Russell
    (Tucson: University of Arizona Press), 639

\pp Landolt, A. U. 1983, AJ, 88, 853

\pp Liou, J.-C. \& Zook, H. A. 1999, AJ, 118, 580

\pp Marcy, G. W., Cochran, W. D., \& Mayor, M. 2000,
    in ``Protostars and Planets IV'',
    ed. V. Mannings, A.P. Boss \& S.S. Russell
    (Tucson: University of Arizona Press), 1285

\pp Metchev S. A. \& Hillenbrand, L. A. 2003, in ``Debris disks and
    the formation of planets'', eds. L. Caroff \& D. Backman,
    ASP Conf. Series

\pp Metchev S. A., Hillenbrand, L. A., \& Meyer, M. R. 2003, ApJ, submitted

\pp Micela, G., Sciortino, S., Harnden, F. R., Kashyap, V., Rosner, R.,
    Prosser, C. F., Damiani, F., Stauffer, J., \& Caillault, J.-P.
    1999, A\& A, 341, 751

\pp Moro-Martin, A., \& Malhotra, R. 2002, AJ, 124, 2305

\pp Ozernoy, L. M., Gorkavyi, N. N., Mather, J. C., \& Taidakova, T. T.
    2000, ApJ, 537, L147

\pp Pantin, E., Lagage, P. O., \& Artymowicz, P. 1997, A\&A, 327, 1123

\pp Perryman, M. A. C. et al. 1997, A\& A, 323, L49

\pp Pollack, J. B., Hollenbach, D., Beckwith, S., Simonelli, D. P.,
    Roush, T., \& Fong, W. 1994, ApJ, 421, 615

\pp Prosser, C. F., Randich, S., Stauffer, J. R., Schmitt, J. H. M. M.,
    \& Simon, T. 1996, AJ, 112, 1570

\pp Quillen, A. C., \& Thorndike, S. 2002, ApJ, 578, L149

\pp Schneider, G., et al.\ 1999, ApJ, 513, L127

\pp Silverstone, M. 2000, UCLA PhD thesis
 
\pp Skrutskie, M. F., Dutkevich, D., Strom, S. E., Edwards, S., 
    Strom, K. M., \& Shure, M. A. 1990, AJ, 99, 1187

\pp Soderblom, D. R., Jones, B. F., Balachandran, S., Stauffer, J. R.,
    Duncan, D. K., Fedele, S. B., \& Hudon, J. D 1993, 105 2299

\pp Spangler, C., Sargent, A. I., Silverstone, M. D., Becklin, E. E.,
    \& Zuckerman, B. 2001, ApJ, 555, 932

\pp Sylvester, R. J., Dunkin, S. K., \& Barlow, M. J. 2001, MNRAS, 
	327, 133

\pp Telesco, C. M., et al.\ 2000, ApJ, 530, 329
 
\pp Tokunaga, A. T. 1988, Infrared Telescope Facility (IRTF) Photometry Manual

\pp Voges, W. et al. 1999, A\& A, 349, 389

\pp Weinberger, A. J., et al.\ 1999, ApJ, 525, L53

\pp Wichmann, R., Schmitt, J. H. M. M., \& Hubrig, S., 2003, A\& A., 399, 983

\pp Wilner, D. J., Holman, M. J., Kuchner, M. J., \& Ho, P. T. P.
    2002, ApJ, 569, L115

\pp Wolf, S., \& Hillenbrand, L. A. 2003, ApJ, in press

\pp Wyatt, M. C., Dent, W. R. F., \& Greaves, J. S. 2003, MNRAS, 342, 876 

\pp Wyatt, M. C., Dermott, S. F., Telesco, C. M., Fisher, R. S., Grogan, K.,
    Holmes, E. K., \& Pina, R. K. 1999, ApJ, 527, 918

\pp Zuckerman, B., \& Becklin, E. E. 1993, ApJ, 414, 793

\clearpage
\begin{table}
\begin{center}
TABLE 1\\
Flux measurements\\
\vskip 2mm
\begin{tabular}{ccc}
\hline\\[-2mm]
$\lambda$ &  Flux & Error \\[-1mm]
(\micron) & (mJy) & (mJy) \\[2mm]
\hline\hline\\[-3mm]
 11.7     & 175 &  10  \\
 17.8     &  85 &   8  \\
 450      & 130 &  $40^{\rm a}$  \\
 850      &  20 &   $4^{\rm a}$  \\[2mm]
\hline\\[-3mm]
\multicolumn{3}{l}{$^{\rm a}$ Calibration uncertainty}
\end{tabular}
\end{center}
\end{table}

\begin{table}
\begin{center}
TABLE 2\\
Properties of spatially resolved sub-millimeter disks around Class V stars\\
\vskip 2mm
\begin{tabular}{lcccccccccl}
\hline\\[-2mm]
Source & SpT & Age & $d$ & $L_\ast$ & $L_d/L_\ast$ & $R_{\rm in}$ & $R_{\rm out}$ & $T_d$ & $M_d$ & Ref. \\
 & & (Myr) & (pc) & ($L_\odot$) & ($10^{-5}$) & (AU) & (AU) & (K) & ($M_\oplus$) & \\[2mm]
\hline\hline\\[-3mm]
HR~4796A       & A0 &    3-10   & 67.1 & 21  & 500 &$\sim 50$& $\sim 200$ &110 & 0.250 & a,b,c,d \\
$\beta$ Pic    & A5 &  10-100   & 19.3 & 8.9 & 200 &$\sim 20$& 210 & 85 & 0.096 & e,f,g,h \\
HD~107146      & G2 &  80-200   & 28.5 & 1.1 & 120 & $> 31$  & 150 & 51 & 0.100 & i \\
Fomalhaut      & A3 & 100-300   &  7.7 & 13  &  10 &    60   & 160 & 40 & 0.018 & e,f,g \\
Vega           & A0 & 150-550   &  7.8 & 60  &   2 &    70   &  90 & 80 & 0.009 & e,f,g,j \\
$\epsilon$ Eri & K2 &$\leq 10^3$&  3.2 & 0.3 &   8 &    30   &  60 & 35 & 0.005 & f,g,k \\[2mm]
\hline\\[-3mm]
\multicolumn{9}{l}{
$^{\rm a}$ Jura et al. (1998),~
$^{\rm b}$ Jayawardhana et al. (1998),}\\
\multicolumn{9}{l}{
$^{\rm c}$ Greaves, Mannings, \& Holland (2000),~
$^{\rm d}$ Schneider, et al.\ (1999),}\\
\multicolumn{9}{l}{
$^{\rm e}$ Holland et al. (1998),~
$^{\rm f}$ Spangler et al. (2001),
$^{\rm g}$ Dent et al. (2000),}\\
\multicolumn{9}{l}{$^{\rm h}$ Lagage \& Pantin (1994); Pantin, Lagage, \& 
Artymowicz (1997),}\\
\multicolumn{9}{l}{
$^{\rm i}$ This work,~ $^{\rm j}$ Wilner et al. (2002),~
$^{\rm k}$ Greaves et al. (1998)}
\end{tabular}
\end{center}
\end{table}

\clearpage
\begin{figure}[ht]
\plotfiddle{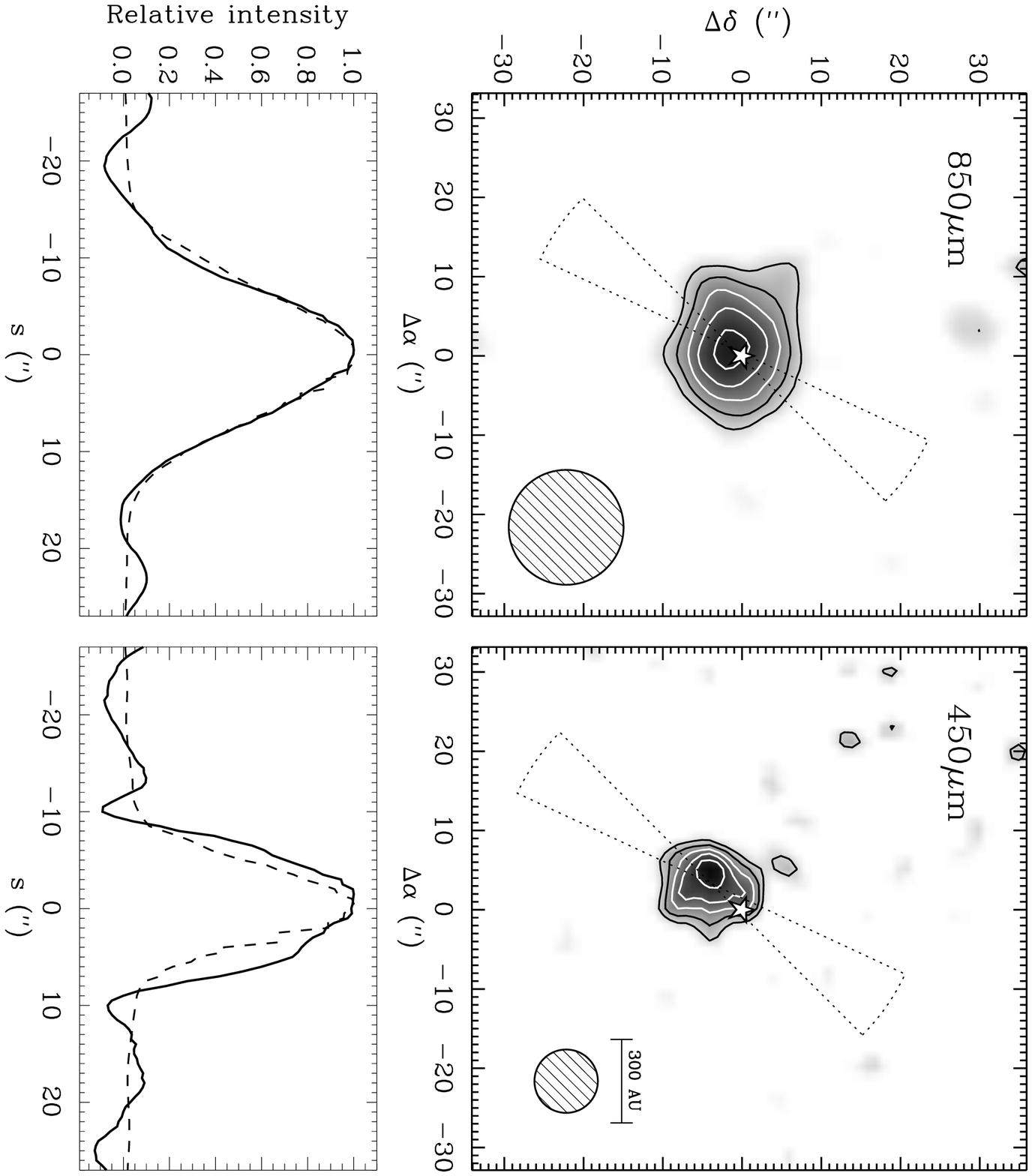}{50pt}{90}{60}{60}{225}{-280}
\vskip 3.1in
\end{figure}
\startfigcap
\noindent{\bf Figure 1:}
SCUBA images of the dust emission around HD~107146.
Coordinates are offset from the J2000 position of the star epoch 2003.13,
$\alpha=12^{\rm h} 19^{\rm m} 6.46^{\rm s}, \delta=16^\circ 32' 53\farcs 4$,
indicated by the star symbol. The left panel shows the emission at 850~\micron.
Contour levels begin at $2\sigma$ and increment by $\sigma$, where
$\sigma=4$~mJy~beam\e\ is the noise in the map.
The greyscale runs from $\sigma$ to $7\sigma$.
The right panel shows the emission at 450~\micron;
contours begin at $3\sigma$ and increment by $\sigma$, where
$\sigma=13$~mJy~beam\e. The greyscale runs from $2\sigma$ to $8\sigma$.
The beam sizes, $14\farcs 5$ at 850~\micron\ and $8''$ at 450~\micron,
are indicated by the hashed circles in the lower right corner of
each figure. Inspection shows that the half power point of the
450~\micron\ image ($\sim 3.5\sigma$) is slightly greater than the beam size.
Averaged cuts of the source and Mars are shown in the
lower panels for each wavelength. The cuts are centered on the peak
of emission and averaged over positions angles $-35^\circ\pm 10^\circ$,
outlined by dotted lines in each image. The Mars profile is shown
as the dashed line and is significantly narrower than HD107146
in the higher resolution 450~\micron\ image.
\endfigcap

\clearpage
\begin{figure}[ht]
\plotfiddle{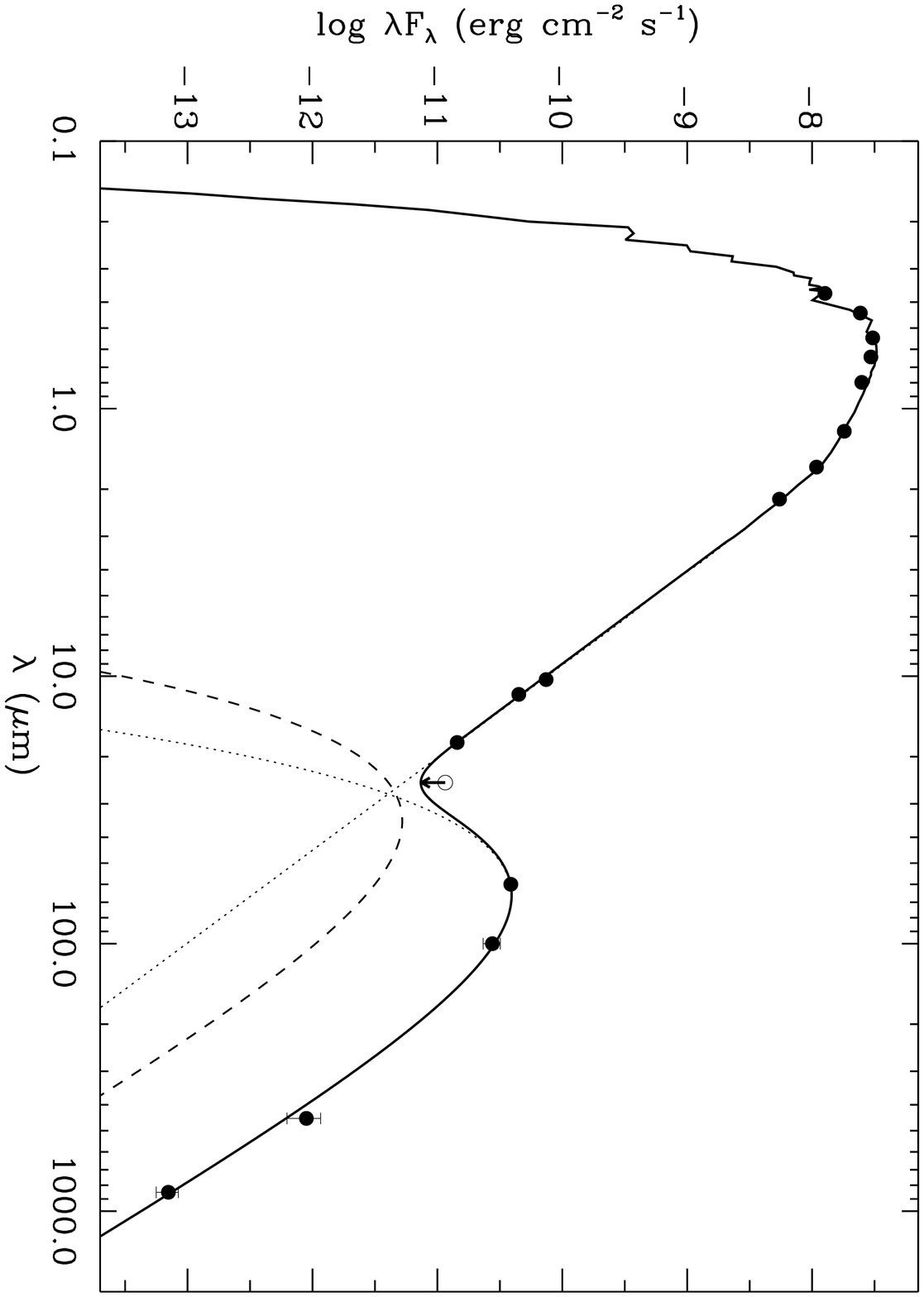}{50pt}{90}{60}{60}{230}{-280}
\end{figure}
\vskip 3.1in
\startfigcap
\noindent{\bf Figure 2:}
Optical to sub-millimeter spectral energy distribution of
HD~107146. The open circle at 25~\micron\ is an upper limit from the IRAS
Faint Source catalog. All other points are detections with error bars
shown when the error exceeds the symbol size.
The double peaked distribution is modeled as the sum of a Kurucz model
of the stellar spectrum and a modified black body fit to the points longward
of 25~\micron\ ($T=51~{\rm K}, M_{\rm d}=0.10~M_\oplus, \beta=0.7$).
The individual contribution of each component is shown by the dotted
lines and the sum of the two by the solid line. The dashed line shows
the maximum allowable $T=100$~K dust component that fits the constraint
of the IRAS 25~\micron\ upper limit. This component is not included in
the overall fit shown by the solid line.
\endfigcap

\end{document}